# Knowledge Management Competence and ISD Vendor Innovativeness in Turbulent Markets

**Research-in-progress**

**Sachithra Lokuge**
School of Accounting, Information Systems and Supply Chain
RMIT University
Melbourne, Australia
Email: ksplokuge@gmail.com

**Maduka Subasinghage**
Business Information Systems Department
Auckland University of Technology
Auckland, New Zealand
Email: maduka.subasinghage@aut.ac.nz

## Abstract

Continuous changes in the technology and the business landscape place high strain on managing knowledge in organisations. Prior researchers highlight a positive connotation with knowledge management competence and organisational innovativeness in a turbulent environment. However, the rapid changes in the market and technology landscape may exert an additional pressure on the employees and such pressures may ultimately hinder organisational innovativeness. Drawing on knowledge management and innovation literature, this research conceptualises a model that investigates this tenacious relationship between knowledge management competence and innovativeness specifically in turbulent dynamic markets, considering information systems development (ISD)-outsourcing as the context. Following a mixed method approach, this research expects to provide guidance for ISD-outsourcing vendors to manage innovation expectations, knowledge management process and performance of the employees in dynamic market conditions.

# 1 Introduction

Globalisation has transformed the socio-economic background of the modern business, where the changing needs of the customers, lifestyles, together with the advancements in technology, have made innovation a necessity for all businesses (Baregheh et al. 2009; Lokuge and Sedera 2018). According to Global Innovation 1000 study, in 2017 total research and development (R&D) spending by the Global Innovation 1000 increased 3.2% to $701.6B (PwC 2017). Such increments in R&D spending in an era where market and technology turbulences are staggering, highlight the importance of innovation. Further, such observations concur with Zahra and Covin (1994), who suggest that innovation is the lifeblood for survival and growth of contemporary corporate organisations.

Prior researchers have highlighted several salient factors that promote innovation in an organisation. They include (i) adequacy of resources including human and financial capital, (ii) conducive long-term organisational learning, (iii) continuous investment strategy and (iv) organisational structures that allow undogmatic organisational thinking (Ettlie et al. 1984; Lokuge et al. 2019; Walther et al. 2015). However, the volatile nature of the markets and the advancement of new technologies mandate organisations to acquire new knowledge and skills to succeed in their business operations. The acquisition of knowledge does not limit to acquiring knowledge regarding market, operational activities and the customers, it also entails the acquisition of knowledge regarding new technologies. As such, concurring with prior research, knowledge management competence can be identified as a salient factor for organisational innovation (Sedera et al. 2003a; Trantopoulos et al. 2017). Considering the importance of acquisition and management of technology related knowledge, managers are exerting pressure on employees to learn and develop new knowledge regarding new and emerging technologies. As per Teece et al. (1997), a turbulent environment catalyses innovativeness of an organisation. As such, prior research alludes to the positive effect of turbulent environment on enhancing the knowledge management competence that leads to innovativeness of an organisation (Keszey 2018; Sedera et al. 2001). While the knowledge management process (phases such as acquisition, retention, sharing and application) of technology related knowledge can drive innovation in an organisation (Alavi and Leidner 2001), the rapid changes in the market and technology landscape may exert an additional pressure on the employees. As such, knowledge management competence may not catalyse innovativeness of the organisation. While prior researchers extol a positive role of knowledge management competence in leading innovativeness of the organisation, it can be argued that high market and technology turbulence may create a negative impact on the innovativeness of the organisation. Such dialectical contradictions regarding the environment turbulences on innovativeness of an organisation motivated us to study this phenomenon. As such, we investigate the overarching research question, *"what is the impact of knowledge management competence on information systems development (ISD)-outsourcing vendor innovativeness, when the market and technology turbulences are high?"*

To investigate this research question, this paper presents a conceptual model. Considering the nature of the research problem, we posit that ISD-outsourcing industry provides an appropriate context for this study. According to Yang (2011), knowledge management competence is important for outsourcing vendors and such practices enable innovation in organisations. As per Weeks and Feeny (2008), there is a strong demand from client organisations for the ISD-outsourcing vendors to innovate beyond the standard agreed specifications of delivery. When the market and technology turbulences are high, the vendor organisations are required to constantly acquire knowledge regarding new technologies and possess knowledge management competence. While prior research alludes to positive role of environmental turbulences to innovative behaviours of organisations, providing service that goes beyond the client's expectations and being innovative can be difficult. As such, this paper will investigate the moderating effect of technology and market turbulence on knowledge management competence and vendor innovativeness contributing to ISD-outsourcing literature.

The remainder of this paper proceeds in the following manner. Next section provides the background of the research followed by the conceptual model and the hypothesis development. Then, the proposed methodology of this study is provided. The conclusion section entails key predictions, contributions to academia and practice. Finally, the appendix provides the survey instrument developed to assess the conceptual model.

# 2 Background

Prior researchers extol the role of knowledge management competence for successful ISD-outsourcing engagements (Gottschalk 2006; Yang 2011). According to Correia & Aguiar (2009, p.1) "knowledge plays a key role in software development, and the effectiveness of how it is captured into artefacts and acquired by other team members, is crucial for project's success." As per Goldoni and Oliveira (2010), ISD-outsourcing activities require constant update of new knowledge regarding the customers, their business processes, industry practices etc. Similarly, prior research (e.g., López-Nicolás and Meroño-Cerdán 2011) highlight the vital role of knowledge management[1] practices that leads to innovation in organisations. Similarly, vendor organisations apply knowledge management practice as a tool for innovation that provides them with competitive advantage over the rivalry organisations (Yang 2011). In recent times with constant market change and constant technology advancements, the intensity of knowledge management competence has become higher. As a result, vendors are not only required to learn about the customer related knowledge, but also new technologies, practices and application of such technologies in business processes (Subasingha et al. 2012). As per Matook and Blasiak (2020), failure to learn new knowledge successfully intensifies the likelihood of errors in application of knowledge, which results in increasing the costs and impeding customer satisfaction. As such, many organisations embrace learning new technologies to enhance customer satisfaction.

While there are constant changes in the market and technology landscape, vendor organisations are constantly required to develop new knowledge to innovate. However, such instances require investment of intensive amount of time, effort, and money. As such, constant learning is challenging. Further, as per Becker (2010) employees are required to abandon bygone knowledge before or during the learning of new technology knowledge. However, such practices are not common in organisations and no proper trainings for employees to manage such vast volatile knowledge. As such, while prior researchers highlight the positive role of knowledge management competence leading to innovativeness in a turbulent environment, it can also be argued that, the constant changes in the technology and market may hinder such abilities. As such, it can be expected that knowledge management competence leading to vendor innovativeness might have a negative connotation when the market and technology turbulences are high.

## 2.1 Proposed Model and Hypothesis Development

Figure 1 conceptualises the relationship between knowledge management competence and vendor innovativeness of ISD-outsourcing engagements.

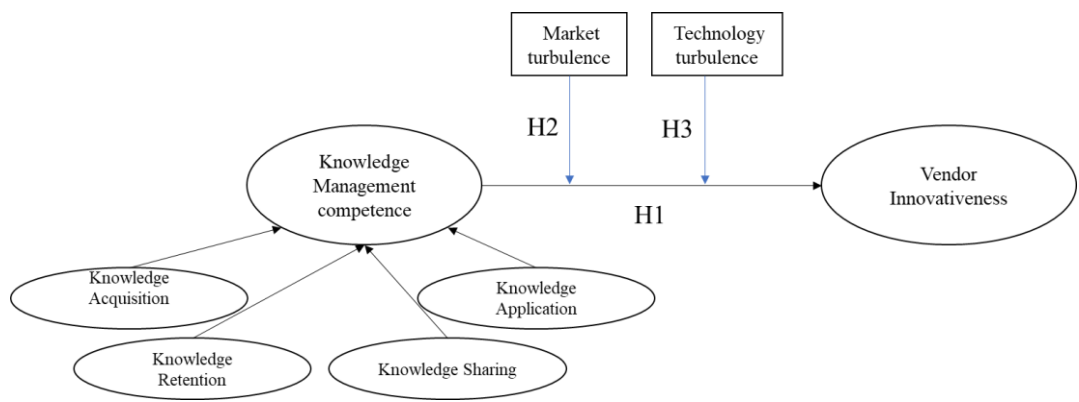

*Figure 1: The conceptual model*

Prior research alludes to positive association between knowledge management and organisational innovativeness (Chen et al. 2010). As per Teece et al. (1997), the need for knowledge management competence is particularly high in an environment where the market and technology turbulences are high. In such environments – like the modern ISD-outsourcing context - organisations are under immense pressure to acquire new knowledge, retain and share knowledge particularly regarding new technologies and opportunities that such technologies have presented to client organisations. When the technology turbulence is high, organisations are required to adopt to new technical knowledge to attain 'first-mover' benefits (Keszey 2018). However, given the dearth and demand for technology skills in turbulent markets (Lichtenthaler 2009), ISD firms will have the challenge of rapid knowledge acquisition, retention, sharing and application. Further, when the market turbulence is also high, ISD-outsourcing vendor organisations are required to provide dynamic services and products to retain customers. Moreover, such turbulences rapidly make current products/services outmoded (Schweitzer

---

[1] According to Yang (2011), knowledge management is the process of identifying, processing, and sharing organizational knowledge to explore new opportunities that improve organizational performance.

et al. 2011). According to Jansen et al. (2006) such market and technology turbulences will have a positive effect on organisations, where they become more innovative. As per Teece et al. (1997), when the market and technology conditions are stable, there is less eagerness to innovate and organisations rely on the existing technologies rather than learning and adopting new technologies. Considering the prior research, we posit the following hypotheses:

H1: *Knowledge management competence is positively associated with vendor innovativeness*

H2: *Market turbulence positively moderate the relationship between knowledge management and vendor innovativeness*

H3: *Technology turbulence positively moderate the relationship between knowledge management and vendor innovativeness*

As per Alavi and Leidner (2001), knowledge management phases (i.e. acquisition, sharing, retention, and application) are distinct and each phase contributes to overall knowledge management competence in the organisation. The construct 'knowledge management competence' was developed following the theoretical guidance of Sedera and Gable (2010) where the phases are conceived as dimensions forming knowledge management competence construct. As such, knowledge management competence is conceived and operationalised as a hierarchical, multidimensional, formative index. As per Hinkin (1995), a good formative construct should collectively represent all the relevant aspects of the variable of interest. Further, in a formative construct, the a-priori model sub-constructs must not co-vary, should not be interchangeable, should cause the core-construct as opposed to being caused by it, and should have different antecedents and consequences in potentially quite different nomological nets (Cenfetelli and Bassellier 2009). Moreover, use of formative constructs in this case provide a 'specific and actionable attributes' of a concept (Mathieson et al. 2001), which is particularly interesting from a practical viewpoint as the weight of the construct can be used to draw practical implications on the importance of specific details and therefore guide practical enforcement on the characteristics. Further, according to Burton-Jones and Straub (2006), a common approach to identify a-priori sub-constructs (and items) is to select from the existing literature. Through acquisition, retention, sharing and application of knowledge regarding new technologies, organisations can enhance their innovation capabilities (Sedera and Lokuge 2019; Sedera et al. 2014; Walther et al. 2018). For example, by acquiring knowledge about new technologies, vendors can improve business processes, reduce cost of business processes and become market leaders in providing new products and services to clients (Lokuge and Sedera 2014).

## 3   Proposed Research Methodology

This research follows a mixed-method approach where we plan to execute a qualitative phase followed by a quantitative phase. The aim of the qualitative phase is to confirm the sub-constructs and the items derived through the literature review. The initial survey instrument derived through literature is provided in Appendix 1. During quantitative phase, content validity of the updated survey instrument will be established by sharing the survey with an expert panel (i.e. 10 employees of ISD-outsourcing vendor organisations) (Sedera et al. 2003b). Thereafter, a pilot test will be conducted and update the instrument accordingly.

The final survey will be shared with employees of ISD-outsourcing vendors. A sample of vendors in ISD-outsourcing was appropriate for the study objectives, as these personnel would be able to comment knowledgeably on behalf of the organisation. Since partial least squares (PLS) structural equation modelling (SEM) method can evaluate complex predictive models, and mapping of formative observed variables (Henseler and Sarstedt 2013), we will employ guidelines of Benitez et al. (2018) for the assessment of the model. To establish the model and construct validation, we will conduct content validity, confirmatory composite analysis, construct validity, testing the measurement model, and observing the additivity and nomological net test. In addition, we will conduct a post-hoc analysis to see how the relationship between knowledge management competence and vendor innovativeness changes when the market and technology turbulence varies.

## 4   Conclusion

The advancement and the proliferation of new technologies along with dynamic market conditions are mandating the ISD-outsourcing vendor organisations to utilize and adopt new technologies in their business endeavours. The continuous technology changes and the market dynamism have exerted an additional pressure on ISD-outsourcing vendors to acquire, retain, share and apply new knowledge regarding new technologies. As per knowledge management literature (Keszey 2018; Storey and Kahn 2010), when the turbulence is high, organisations tend to be innovative (Teece et al. 1997). However, when the technology and market turbulences are high, there is too much pressure exerted on employees, which might negatively affect the innovativeness of the vendor organisation. As such, the objective of this research is to better understand the impact of environment turbulence on knowledge management competence that leads to ISD-outsourcing vendor innovativeness. The study follows a mixed method approach to investigate this phenomenon. Further, by conducting a post-hoc analysis, the study attempts to provide empirical evidences on managing expectations of ISD-outsourcing organisations.

The successful completion of the study will contribute to outsourcing and knowledge management research. We expect that the study findings will help ISD-outsourcing vendors to manage expectations regarding the innovation process, knowledge management process and performance of the employees. This study will provide guidance for managers to drive innovation in turbulent environments. Especially, ISD-outsourcing vendors tend to pressurize their employees to learn new technologies and innovate in a frequent manner. Even though this is the common understanding among managers, this process may hinder innovativeness of the organisation. The qualitative phase of this research is still in progress at the time of submission of this paper. As discussed in the proposed methodology section, this phase involves data collection through interviews to confirm the sub-constructs and the items of the survey instrument developed through the literature review. A survey will be distributed in the quantitative phase conducting SEM-PLS analysis and post-hoc analyses.

# Appendix 1

| Construct | Items | Reference |
|---|---|---|
| Knowledge Retention | 1. My organisation is required to devote substantial resources to maintain KM systems to keep up with constant technology changes.<br>2. My organisation is required to devote substantial resources to maintain technology tools (collaborative tools, knowledge bases, searching tools, document management systems, intelligent systems, etc.) to facilitate knowledge storage.<br>3. My organisation is required to devote substantial resources to retain employees. | (Lin et al. 2016) |
| Knowledge Sharing | 1. My organisation is required to devote substantial resources to share knowledge effectively to keep up with constant technology changes.<br>2. My organisation is required to devote substantial resources for training programs to keep up with constant technology changes. | Developed |
| Knowledge Acquisition | 1. My organisation is required to devote substantial resources to gain new knowledge of different technologies important for forthcoming projects.<br>2. My organisation is required to devote substantial resources to gain new knowledge of specific technologies important for forthcoming projects.<br>3. My organisation is required to devote substantial resources to gain new knowledge of technologies that are central to forthcoming projects.<br>4. My organisation is required to devote substantial resources to gain new hands-on experiences with technologies that are important for forthcoming projects. | (Sullivan and Marvel 2011) |
| Knowledge Application | 1. Overall, my organisation has used new knowledge regarding different technologies effectively and efficiently.<br>2. Overall, my organisation has used new knowledge regarding specific technologies effectively and efficiently.<br>3. Overall, my organisation has used new knowledge regarding central technologies effectively and efficiently. | (Alavi and Leidner 2001; Sedera and Gable 2010) |
| Market Turbulence | 1. Customers in our markets are receptive to new product ideas.<br>2. In our markets, customers' preferences change relatively fast.<br>3. New customers tend to have product-related needs that are different from those of existing customers.<br>4. We mainly address the same customer base that we did in the past. | (Lichtenthaler 2009) |
| Technology Turbulence | 1. The technology in our markets is changing rapidly.<br>2. Technological developments in our markets are rather minor.<br>3. Technological changes provide big opportunities in our markets.<br>4. It is very difficult to forecast where the technologies in our markets will be in the next five years.<br>5. A large number of new products in our markets have been made possible through technological breakthroughs. | (Lichtenthaler 2009) |
| Vendor Innovativeness | 1. My organisation readily accepts innovations based on research results.<br>2. Management in my organisation actively seeks innovative ideas.<br>3. Innovation is readily accepted in this organisation.<br>4. People are penalized for new ideas that don't work.<br>5. Innovation in this organisation is perceived as too risky and is resisted. | (Venkatesh and Bala 2012) |

# Copyright